\documentclass[usenatbib,conference]{basi}
\usepackage[british]{babel}
\usepackage[varg]{txfonts}
\usepackage{rotating}
\usepackage{dcolumn}

\begin{document}

\title[Spectroscopic charaterization of FHLC stars  
and a newly found HdC star]{Spectroscopic charaterization of FHLC stars from 
the Hamburg/ESO survey and a newly found HdC star}

\author[Aruna Goswami ]%
       {Aruna Goswami\thanks{e-mail:aruna@iiap.res.in}   \\
        Indian Institute of Astrophysics, Bangalore 560 034}

\pubyear{2011}
\volume{00}
\pagerange{\pageref{firstpage}--\pageref{lastpage}}
\date{Received \today}
\maketitle
\label{firstpage}
\begin{abstract}
 The sample of candidate faint high latitude carbon (FHLC) stars chosen from 
the Hamburg/ESO survey is a potential source to search for objects of rare 
types. From medium resolution spectral analyses of about 250 objects from 
this sample, the object HE~1015$-$2050, was found to be a hydrogen-deficient 
carbon (HdC) star. Apart from U Aquarii, HE 1015-2050 is the only example, 
till now,  of a Galactic cool HdC star that is characterized by strong 
spectral features of light s-process element Sr, and weak features of heavy 
s-process elements such as Ba. This object, with its enhanced carbon and 
hydrogen-deficiency, together with anomalous s-process spectral features, 
poses a challenge as far as the understanding of its formation mechanism 
is concerned. We discuss possible mechanisms for its formation in the 
framework of existing scenarios  of HdC star  formation.

\end{abstract}

\section {Introduction}
The Hamburg/ESO survey was initiated in the year 1990, for 
the Southern Hemisphere to complement the Hamburg/Quasar survey (HQS) that 
covers the Northern sky except the Galactic plane  (Wisotzki et al. 1996, 
Reimers \& Wisotzki 1997, Wisotzki et al. 2000). This survey was based on 
digitized objective-prism photographs taken with ESO 1m Schmidt telescope. 
An area of 9500 deg$^{2}$ in the southern sky was covered with an average 
limiting magnitude  B$\sim$ 17.5 mag, on the prism plate. 
Christlieb et al. (2001) have used the Hamburg/ESO survey (HES) to 
augment the number of known FHLC stars. The HES spectra cover a  wavelength 
range  of 3200 - 5200 \AA\, and the seeing limited  spectral
resolution  is  typically 15 \AA\, at H$_{\gamma}$.  An automated procedure 
based on the detection of C$_{2}$ and CN molecular bands of the spectra 
was used to identify  the carbon stars. This procedure resulted
 403 candidate faint high latitude carbon stars from a set 
of 329 plates  that cover an area of 6400 deg$^{2}$ (87\% of the survey area)  
to the   magnitude limit V ${\sim}$ 16.5. The surface density of FHLC stars 
$\sim$ 0.072 $\pm$ 0.05 deg$^{-2}$, is  about 2-4 times higher than those 
obtained from  previous objective prism and CCD surveys at high Galactic 
latitude (Sanduleak \& Pesch 1988; MacAlpine \& Lewis 1978, Green et al. 1994).
However, it is to be  noted that the  majority of the carbon stars known today 
mostly come from the Sloan Digitized Sky Surveys (SDSS). 

Although, at high Galactic latitude the surface density of FHLC stars is low, 
different classes  of carbon stars  populate the region: a) the normal AGB 
stars, carbon-enriched by dredge-up during the post main-sequence phase, 
which are found among the N-type carbon stars, b) FHLC stars showing 
significant proper motions and having  luminosities of  main-sequence 
dwarfs, called dwarf carbon stars (dcs) and c) the  CH giants, similar to 
the metal-poor carbon stars found in Globular clusters and some dwarf 
spheroidal  galaxies. Among these, warm carbon stars of  C-R type  
are also likely to be present. The  sample of stars listed by Cristlieb 
et al. being high latitude objects  contains a mixture of these objects. 
Using  medium resolution spectral analysis we have  classified these  
objects  based on their spectral chatacteristics. The spectra were obtained  
using the Himalayan Faint Object Spectrograph Camera (HFOSC) attached to 
the Himalayan Chandra Telescope (HCT) at the Indian Astronomical
Observatory (IAO), Hanle during  2005 - 2010. The grism and the camera
combination used for observations provided a spectral resolution of
$\sim$ 1330($\lambda/\delta\lambda$); the observed bandpass is about 3800 -
6800 \AA\,. The spectra of a  few objects were also acquired using the OMR
spectrograph at the cassegrain focus of the 2-3-m Vainu Bappu Telescope
(VBT) at Kavalur. With a 600 line mm$^{-1}$ grating, we get a dispersion of
2.6 \AA\, pixel$^{-1}$. The wavelength range covered is  4000 - 6100
\AA\, and the resolution   $\sim$ 1000.

The membership of a star in a particular class is established from a
comparison with  the  spectral atlas of carbon stars of Barnbaum et al. 
(1996). Further details on the 
detection procedure  and the spectral characteristics of these objects 
are available in Goswami (2005) and Goswami et al. (2007, 2010a). In our
sample of 403 stars, HE~1015$-$2050 is found to exhibit spectral
charateristics of  HdC stars. Its  photometric parameters,
along with those  of the comparison HdC star of RCB 
type U~Aqr  are given in Table 1. HdC stars are a rare class of objects; 
only five non-variable HdC and fifty five RCB type stars are known so far 
in our Galaxy. The origin of these objects is still debated  and they
are poorly understood due to a lack of statistically significant sample. Each 
addition to this rare group of objects is therefore important.

{\footnotesize
\begin{table}
\centering
{\bf Table 1: Photometric parameters of HE~1015$-$2050 and the \\
comparison  star  U~Aqr }\\
\small
\begin{tabular}{ccccccccc}
\hline
Star No.  &  $l$  & $b$  & B  & V & B-V$^{a}$&J & H & K \\
          &       &      &    &   &          &  &   &    \\
\hline
HE 1015-2050 &  261.31 &  29.08 & 16.9 & 16.3 & 0.67 &  14.977 & 14.778 &  14.504 \\
HE~2200-1652&   39.15 & -49.81 & 12.1 & 11.1 & 0.99  &  9.562 & 9.283 & 8.961 \\
( U Aqr)   &          &        &      &      &       &        &    &          \\
\hline
\end{tabular}

$^{a}$ From Christlieb et al. (2001)\\

\end{table}
}

\section {Spectral   characteristics of HdC stars }
The spectra were classified using the  following spectral characteristics:
a) strength of band-head of the CH band around 4300 \AA\,, 
b) prominence of secondary P-branch head near 4342 \AA\,, 
c)  strength/weakness of the Ca I feature at 4226 \AA\,,  
d) isotopic band strength of C$_{2}$ and CN, in particular the Swan bands 
 of $^{12}$C$^{13}$C and $^{13}$C$^{13}$C near 4700 \AA\,,
e) strength of other C$_{2}$ bands in the 6000$-$6200 \AA\,  region,
f) $^{13}$CN band near 6360 \AA\,  and other CN bands across the 
wavelength range, 
g) strength of s-process elements such as Ba II features at 4554 and 
6496 \AA\,.

The  Hydrogen-deficient supergiants comprise of three sub-classes:
a) Extreme  helium stars (EHes) of spectral types F and G, b) R Coronae 
Borealis  (RCB) stars of spectral types F and G and c) hydrogen-deficient
carbon (HdC) stars that are much cooler than EHes and RCBs. Similarities 
in chemical compositions indicate an evolutionary link between EHes and 
RCBs. Whether there is an evolutionary link  between these stars and HdCs
is not known. The abundance analysis of HdC stars is difficult because
their spectra are dominated by molecular bands.

HdC stars are spectroscopically similar to the RCB stars. However, infra-red
excesses and deep light minima that are characteristics of RCBs are absent
in  HdCs, instead they show small-amplitude light variations. Hydrogen 
deficiency 
and weaker CN bands relative to C$_{2}$ molecular bands are the two primary 
spectral characteristics of RCB stars. The strong C$_{2}$ molecular bands 
are seen in the spectra of cool RCB stars. They are weakly visible in warm 
RCB stars.  As Balmer lines are weak in carbon stars, the strength/weakness 
of CH band in C-rich stars provides a measure  of hydrogen deficiency.
We have used  hydrogen deficiency and the relative strength of C$_{2}$ bands 
in the 6000 - 6200 \AA\, region  and the CN bands near 6206 \AA\, and 
6350 \AA\,   as important classification criteria for HdC stars.

The spectrum of HE~1015-2050 is characterized by strong C$_{2}$ molecular
bands. However, G-band  (CH  around 4310 \AA\,) is only marginally
detected  indicating that HE~1015-2050 is a hydrogen-deficient carbon star.
 HE~1015$-$2050 bears a remarkable similarity with 
 U~Aqr, a cool RCB star. Similar to  U~Aqr, it
  exhibits an anomalously strong feature of Sr II at 
4077 \AA\,. Y II line at 3950 \AA\, is also clearly detected  in both  
 spectra and no significant enhancement of Ba II features at 
4554 \AA\, and 6496 \AA\, is seen. Fe I feature at 4045 \AA\, is clearly 
detected.
The feature due to Na I D also appears to be strong. The H$_{\alpha}$ feature 
is not detected. The most striking feature in HE~1015$-$2050 and U~Aqr 
is the Sr II at 4215 \AA\,, this feature is blended with the nearby 
strong blue-degraded (0,1) CN 4216 band head in HD~182040 and ES Aql 
(Fig. 1).

 RCB stars are also characterized by their   location in 
the J-H and H-K planes with respect to cool carbon stars. The Two Micron 
All Sky Survey (2MASS) measurements (Skrutskie et al. 2006) place 
HE~1015$-$2050 on the J-H versus H-K plane along with the cool LMC
RCB stars  supporting our classification.

\begin{figure}
\centerline{\includegraphics[width=10cm, height=8.0cm]{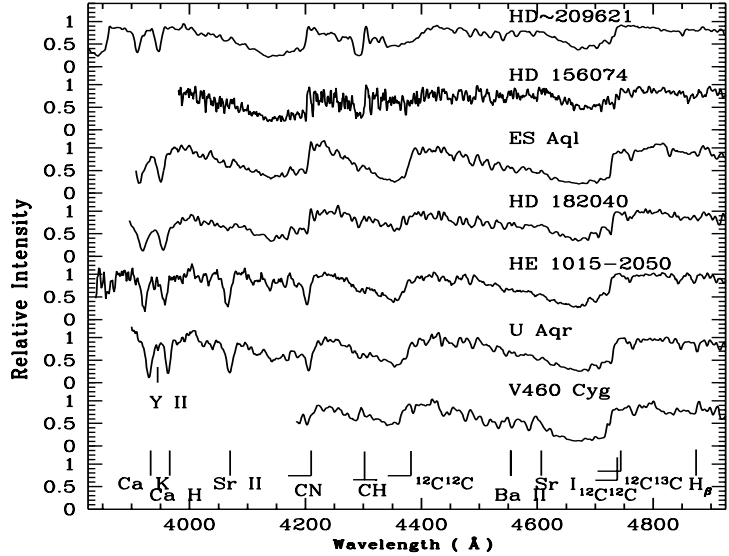}}
\caption{ A comparison between the spectrum of  HE 1015-2050 with the spectra
of V460 Cyg (C-N star),  U Aqr, ES Aql (cool HdC stars of RCB type),
HD~182040 (a non-variable HdC star),  HD~156074 (C-R star), and  HD~209621 
(CH star)  in the wavelength region 3850-4950 \AA\,. G-band of CH distinctly 
seen in the CH and C-R star's spectra are barely detectable in the spectra
of HE 1015-2050 and other HdC stars spectra.
 The large enhancement of Sr II at 4077 \AA\,  in the spectrum of U Aqr
is easily seen to appear with almost equal strength in the spectrum of
HE 1015-2050.  Y II line at 3950 \AA\, is  prominent in the
spectrum of HE~1015-2050,  this line of Y II is also considerably
strengthened in U Aqr. These two features of Sr II and Y II  are not 
observed in the spectra of HD~182040 and ES Aql. The  spectrum of 
HE~1015-2050 compares closest to the spectrum  of the  HdC star U Aqr of 
RCB type (Goswami et al. 2010b).}
 
\end{figure}

\begin{figure}
\centerline{\includegraphics[width=10cm, height=8.0cm]{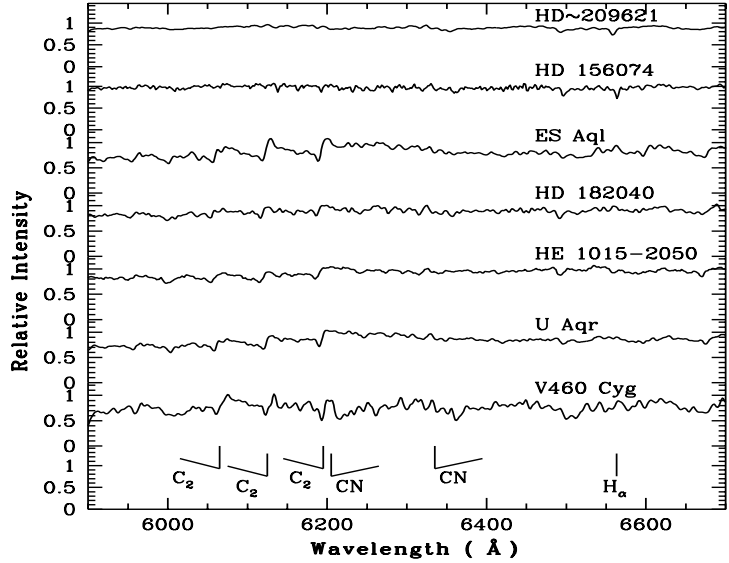}}
\caption{ Same as Fig. 1, except for the wavelength region 5900-6700 \AA\,.
The CN bands which appear with almost equal strengths in the  spectrum of
the CN star V460 Cyg is almost absent (or barely detectable) in the
spectra of HE~1015-2050 and the HdC star U Aqr of RCB type.
H$_{\alpha}$  feature  is distinctly  seen in the spectra of the CH and C-R
star HD~209621 and Hd~ 156074 respectively. This feature is not detectable
in the spectra of  HE~1015-2050 and HdC stars. Non detection of H$_{\alpha}$,
and marginal detection of G-band of CH  (Fig. 1) hints at hydrogen-poor 
nature of the object (Goswami et al. 2010b). }
 
\end{figure}

\section{ Discussion and Conclusions}

The strong Sr II features observed in HE~1015$-$2050 are  indications of
enhanced s-process abundances.  In late type stars such as CH giants and 
carbon-enhanced metal-poor objects, the observed enhanced abundances
of s-process elements   are generally explained on the basis of a binary 
picture 
in which the primary companion low-mass object while evolving through the
Asymptotic Giant Branch (AGB) stage transfers  the s-enhanced material to
the secondary companion (McClure 1983, 1984; McClure \& Woodsworth 1990). 
In case of HE~1015$-$2050 such an explanation is presently not
 applicable  as its binary status is not yet known. 
Although  none of the RCB and HdC stars  known so far, 
are known  to be binary, long-term radial
velocity monitoring of HE~1015$-$2050 would be useful to know its binarity.

The stellar atmosphere of HE~1015$-$2050 has an estimated effective 
temperature (T$_{eff}$) of 5263 K, derived using semi-empirical temperature
calibration relations from  Alonso et al. (1996). This temperature 
estimate is very 
similar to those of cool Galactic RCB stars  such as S~Aps, WX~CrA,
and U~Aqr (T$_{eff}$ $\sim$ 5000 K, Lawson et al. 1990).
Although the resolution of our spectrum is not adequate to derive quantitative 
estimates 
 one could expect  this object to have similar s-elements abundance
as that of U~Aqr, as they exhibit very similar spectra.
Vanture et al. (1999) have determined  abundances  of  Rb, Sr, Y, and Zr
in U~Aqr that  are  found to be greatly enhanced
 but Ba did not show any significant enhancement. Estimates
of Vanture ([Y/Fe] = $+$3.3, [Zr/Fe] = $+$3.0, and [Ba/Fe] = $+$2.1)
are in general agreement with those of Malaney (1985) but larger than
the estimates of Bond (1979). 

The abundances of light s-process elements  in solar material is attributed 
to `weak s-process'  described by a 
single-neutron irradiation (Beer \& Macklin 1989).  
The main component of s-process occurs through  partial mixing of protons 
into the radiative $^{12}$C layer during thermal 
pulses that initiate  the chain of reactions 
$^{12}C(p,\gamma)^{13}N(\beta)^{13}C(\alpha,n)^{16}O$,
in a narrow mass region of the He intershell called the $^{13}C$ pocket.
The reaction  $^{13}C(\alpha,n)^{16}O$ acts as the source of neutrons 
 (Iben \& Renzini 1982, Lattanzio 1987) in this process. The `weak s-process'
is assumed to occur in massive stars in He or C-burning phase.  
Although,  no observational support exists for this scenario,
$^{22}Ne(\alpha,n)^{25}Mg$  reaction is  believed to be the source of neutrons. 
  This reaction  has  limited efficiency as most of the 
neutrons liberated
are  absorbed by light nuclei, and  a few remain  available for Fe-seed
nuclei to capture.  This process allows production of light s-process
nuclei with mass numbers 65 $<$ A $<$ 90 (Prantzos et al. 1990) and
 Sr with mass number A = 87 falls within this range. Weak s-process could
therefore be a likely mechanism responsible for the observed Sr in U~Aqr 
and  HE~1015$-$2050.

Main-sequence objects with strong Sr are quite common but they also
show equally strong Ba. However, in the globular cluster $\omega$ Cen 
Stanford et al. (2006) have noted one object 2015448 with anomalously
strong Sr and weak Ba. The  effective temperature, 
 surface gravity and the metallicity ([Fe/H])   of this 
object are respectively 5820 K, 4.2 and $-$0.7. The Sr and Ba 
abundances with respect to Fe are   respectively [Sr/Fe] = +1.6 
and [Ba/Fe] = 0.6 (Stanford et al. 2006). A similar mechanism responsible 
for 2015448 formation may also hold good for HE~1015$-$2050 and needs further
investigation. We note, however, a primary difference between 2015448 and 
HE~1015$-$2050;
while the former shows carbon depletion ([C/Fe] = $-$0.5),
HE~1015$-$2050 shows  enhancement of carbon.   

Low mass hydrogen-deficient stars are associated with late stage of
stellar evolution and are believed to be in a  short-lived evolutionary
phase. The characteristic light decline of five or more magnitudes shown
by RCB stars  within  a few days  from the onset of minimum
(followed by slow recovery to maximum light) is suggested   to be
due to directed mass-ejections, which  is believed to be  primarily a 
signature of surface activity rather than chemical peculiarity. 
Hydrogen-deficient stars that do not exhibit such irregular fadings may 
indicate absence of such mass-ejection episodes; but, whether that is 
characteristic of a particular evolutionary stage is not yet established. 
Detailed  spectroscopic  as well as  photometric studies could  provide  
insight  into these aspects. In addition, extended photometric observations  
of HE~1015$-$2050  would be useful to detect short as well as  long term 
photometric variations observed in RCBs.

{\it Acknowlegment} \\
AG gratefully acknowleges the LOC for local hospitality during the workshop
and thanks Sunetra Giridhar for her comments and suggestions on the manuscript.

\end{document}